\documentclass[12pt,epsf]{article}
\usepackage{tabularx}
\usepackage{amssymb,amsmath,amscd}
\usepackage{amsfonts}
\usepackage{epsfig,float,wrapfig}
\newcommand{\nc}{\newcommand}
\def\frac#1#2{{\textstyle {#1 \over #2}}}

\def\d{{\rm d}}

\nc{\beq}{\begin{equation}}
\nc{\eeq}{\end{equation}}
\nc{\beqa}{\begin{eqnarray}}
\nc{\eeqa}{\end{eqnarray}}
\nc{\lsim}{\begin{array}{c}\,\sim\vspace{-21pt}\\< \end{array}}
\nc{\gsim}{\begin{array}{c}\sim\vspace{-21pt}\\> \end{array}}
\nc{\appA}{}
\nc{\appB}{}
\nc{\appC}{}
\nc{\appD}{}
\nc{\appE}{}
\nc{\Eqr}[1]{(\ref{#1})}
\nc{\mysection}[1]{\setcounter{equation}{0}\section{#1}}

\nc{\myappendix}[1]{\section*{#1}\setcounter{equation}{0}}

\def\dk{d\bar{k}}

\def\bpsi{\bar\psi}

\def\bta{\bar\eta}

\def\&{and}

\def\dS{\partial\!\!\!/}
\def\kS{k\!\!\!/}
\def\qS{q\!\!\!/}
\def\pS{p\!\!\!/}

\def\epsilonS{\epsilon\!\!\!/}
\def\d{\partial_\mu} 
\def\A { A_\mu (x) }
\def\AS {A\!\!\!/}

\def\bu{ \bar{u} }

\def\D{ {\cal D} }
\def \Eslash {E \kern-.5em\slash}

\def\hpa#1#2#3{           {\it Helv. Phys. Acta  }{\bf #1}, #2 (#3)} 

\def\imp#1#2#3{           {\it Int. J. Mod. Phys. }{\bf #1}, #2 (#3)}

\def\nc#1#2#3{           {\it Nuovo Cim.  }{\bf #1}, #2 (#3)}

\def\np#1#2#3{           {\it Nucl. Phys. }{\bf #1}, #2 (#3)}
\def\pl#1#2#3{           {\it Phys. Lett. }{\bf #1}, #2 (#3)}

\def\pr#1#2#3{           {\it Phys. Rev. }{\bf #1}, #2 (#3)}
\def\prep#1#2#3{         {\it Phys. Rep. }{\bf #1}, #2 (#3)}

\def\rmp#1#2#3{          {\it Rev. Mod. Phys. }{\bf #1}, #2 (#3)}

\begin{document}
\begin{titlepage}
\renewcommand{\thefootnote}{\fnsymbol{footnote}}
\begin{center}
\hfill
\vskip 1 cm
{\large \bf Path integral regularization of QED by means \\
of Stueckelberg fields}
\vskip 1 cm
{
  {\bf J. L. Jacquot}\footnote{jacquot@lpt1.u-strasbg.fr}
   \vskip 0.3 cm
   {\it Laboratoire de Physique Th\'eorique,
        3 rue de l'Universit\'e,
        67084 Strasbourg Cedex, FRANCE}\\ }
  \vskip 0.3 cm
\end{center}

\vskip .5 in
\begin{abstract}
With the help of a Stueckelberg field we construct a regularized $U(1)$ gauge invariant action through the introduction of cutoff functions.
This action has the property that it converges formally to the unregularized action of QED when the ultraviolet cutoff goes to infinity.
Integrating out the Stueckelberg field  exactly we obtain a  regularized action, which describes the interaction of the photon with a massive screened fermion.
This action is fully gauge invariant and the relation to standard QED is confirmed at the tree level and at the one loop order.
\end{abstract}

\end{titlepage}
\renewcommand{\thepage}{\arabic{page}}
\setcounter{page}{1}
\setcounter{footnote}{0}
\mysection{Introduction}
Fifty years ago, the pioneer work of Stueckelberg \cite{STUE1} on the ``normalization group'' paved the way  to our modern understanding of the 
concept of renormalization in Quantum Field Theory.
Nowadays it is believed that the  laws of nature can be described  at different scales through effective actions which are obtained by integrating out  the low or high energy modes of the fields  or even some field completely.
The evolution of the effective actions with the scale are then governed by approximate or exact renormalization  group equations (RGE) \cite{BAGN,LITI}.

The RGE can be obtained  exactly if the symmetries of the underlying theory are preserved by the new effective action. 
In practice, integrating over energy modes of the fields, supposes that some regularization scheme is understood.
In general it is hard to preserve  the symmetry of the original action  by a continuous nonperturbative regularization if this symmetry  is not a trivial one.
This is the case for gauge invariance, because the gauge transformation mixes the low and  high energy modes of the fields.
In this respect it was shown  at the one loop order \cite{LIAO} that  the operator cutoff regularization is in fact suited for the study of non abelian gauge theories with the Wilson-Kadanoff renormalization group methods.
The deduction of the exact RGE without enlarging the gauge group  and without a need of preregularization \cite{MORR1}, or without loss of covariance \cite{SIMI2}  is still lacking for non abelian gauge theories.
Even in the case of QED the Wilson-Kadanoff renormalization group was applied only for a very specific choice of the cutoff functions \cite{SIMI1}.
These cutoff functions corresponding to a mass term for the photon and the electron were chosen in such a way that the breaking of gauge invariance be only confined at the tree level and not  in the loop expansion.
However  a gauge invariant regularized action was constructed explicitly in \cite{JLJ} by smearing the pointlike interactions of the fields by arbitrary cutoff functions.
In this scheme, the action contains an infinite sum of vertices in order to restore the gauge invariance  and  a gauge invariant counterterm, quadratic in the cutoff,  was added by hand to the action in order to ensure a vanishing renormalized photon mass.
Due to the complexity of the action higher order loop calculations are not easy to perform analytically.

In the present  work we construct an exact gauge invariant and regularized action with the help of quite arbitrary cutoff functions.
The path integral of the theory is thus regularized in a gauge invariant manner.
This action is formally the one described in \cite{JLJ}, but gauge invariance is implemented by the addition of a Stueckelberg \cite{STUE2,RUEG} field.
Since there is no constraint on the scalar gauge function, after a suited unitary gauge transformation we can  integrate out the Stueckelberg field.
The  effective action thus obtained is gauge invariant and its  form is well suited for the application of exact RGE techniques.
If compared to standard QED it contains an additional four Fermi interaction  in addition to the  gauge invariant counterterm  needed to cancel the quadratic divergence which arises at the one loop order in  the polarization operator \cite{JLJ}.
These two supplementary terms  ensure  that the standard predictions of QED are recovered at  the tree level and at the one loop order. 
Though this new regularized action is suited for nonperturbative analysis, we derive the regularized Dyson-Schwinger (DS) equations to show that the one particle irreducible (1PI) function are similar to that of QED at the one loop order.
\mysection{The regularized gauge invariant action }
In order to obtain a gauge invariant regularization of the DS equations of motion of QED, we start with the action
\beq
\label{action1}
S_{Reg}(\psi ,\bpsi ,A,\Phi) = S_{\psi A \Phi} + S_{ A \Phi},
\eeq
where the action
\beqa
\label{actionpsia} 
\ S_{\psi A \Phi}  &=&\int dxdzdz'~ \bpsi (z)~e^{ieL(z)}\Bigg [\rho_1 
(z,x)\bigg (i\dS\ -m \bigg)
 \rho_1(x,z' )  \nonumber  \\
& & -e\rho_2 (z,x)\bigg ({ \mathbf{\AS}} (x) + (\dS L(x))\bigg  
)\rho_2 
(x,z' 
)\Bigg ]e^{-ieL(z')} \psi (z') \nonumber  \\
& &  - \frac {1}{4} \int dx~  F_{\mu \nu}  F^{\mu \nu },
\eeqa
is formally similar to the action given in \cite{JLJ} and where $S_{ A \Phi}$ is defined in terms of the Stueckelberg boson field $\Phi$ as
\beqa
\label{actionaphi} 
\ S_{ A \Phi} &=& \frac{1}{2} \int dx~ ( \partial^{\mu} \Phi +M A^{\mu} )(  \partial_{\mu} \Phi + M A_{\mu} ) \nonumber  \\
& &- \frac {1}{2} \int dx~ \left [ \partial^{\mu}  \left (A_{\mu} + \frac{1}{M}\partial_{\mu}  \Phi \right )  \right ]^2  - \frac {1}{2\xi} \int dx~ \left [ \partial^{\mu}  \left (A_{\mu} + \frac{1}{2M}\partial_{\mu}  \Phi \right )  \right ]^2.
\eeqa
In this expression the ultraviolet  (UV) or infrared (IR) regularization can be implemented through the real cutoff scalar functions
\beq
\label{cutf1}
\rho_i (x,y) = \int \dk ~e^{-ik(x-y)}\rho_i (\frac{k^2}{\Lambda^2},\kappa ),
\eeq
$\kappa $ and $\Lambda$ being respectively the IR and UV cutoff scales.
These functions are supposed to be  regularized forms of Dirac's $\delta $ function, i.e.
\beq
\label{cutf2}
 \lim_{\Lambda \to \infty,\kappa  \to 0}\rho_i (x,y)=\delta (x-y).  
\eeq
The operator ${ \mathbf{A}}^{\mu}$ is the smeared gauge field 
\beq
\label{sma}
{\mathbf{A}}^{\mu}(x)=\int dy~ \rho_3(x,y)A^{\mu }(y),
\eeq
and the link operator
\beq
\label{link}
L(x)=\int dy~ \rho_3(x,y)\frac{\Phi(y)}{M},
\eeq
is the  smeared Stueckelberg field $\Phi$ divided  by a mass parameter $M$.
If the  cutoff functions are essential to regularize the UV divergencies, as we will see later, the action  (\ref{action1}) can also be regularized in the IR domain only with the help of the mass parameter $M$.
The precise form  of the cutoff  functions  (\ref{cutf1}) are  determined by the kind of exact RGE one wants to deduce.
For instance the UV regularization can be implemented 1) only in the kinetic term through the fermion propagator, or 2) only in the interactions terms, or 3) both in  the kinetic and  interactions terms.
Thus in order to render all Feynman diagrams finite in Euclidean space the product $\rho_1^{-1}(k)\rho_2(k)$  must be a rapid decreasing function of the squared momentum in the UV domain and   in this domain the positive function $\rho_3(k)$ must be at least bounded by a constant.

Notice that 1) if one takes  first the limit $\Lambda \to + \infty$ and $\kappa \to 0$ the contribution (\ref{actionpsia}) to  the action (\ref{action1}) converges  formally to the gauge invariant part of the known  unregularized action of QED.
Then 2) if one  performs the  exact integration on the bosonic field  $\Phi$  after the gauge transformation 
\beqa
\label {gaugetrans0}
\ \A ~&\rightarrow &~\A -\frac{1}{M}\d \Phi(x) \nonumber \\
\ \psi(x)~&\rightarrow &~e^{ie \frac{\Phi(x)}{M} }\psi(x),
\eeqa
the remaining part (\ref{actionaphi}) of the original action (\ref{action1}) gives rise to the  standard gauge fixing term and to the photon mass term $\frac{ M^2}{2} \int dx A_{\mu} A^{\mu}$.
The unregularized action  formally obtained from (\ref{action1}) is thus   the  known unregularized massive action of QED.
Gauge invariance of the bare action implies that in any regularization scheme the longitudinal degrees of freedom of the gauge field, and hence the photon mass, cannot be  renormalized by higher order radiative corrections \cite{ZINN}.
In perturbative  regularization like for instance dimensional regularization \cite{THOOFT}, or Pauli-Villars regularization  \cite{PAULIVI}, the potentially quadratic UV divergences which can occur in the polarization operator of massive standard QED are avoided by construction.
Therefore,  the value of the bare  photon mass in these schemes must be set equal to  zero in order to get a vanishing   renormalized photon mass.
On the contrary, in the regularization scheme under consideration, gauge invariant quadratic UV divergences will occur in the polarization operator  \cite{JLJ}.
As a result, in order to keep the renormalized photon field massless,  the  value of the bare  photon mass must be a divergent quantity as it is usual in Quantum Field Theory.
This value will be fixed in such a way that the gauge invariant part of the photon mass term  will  play the role of a  counterterm. 

Since the asymptotic property 1) of the action  (\ref{actionpsia}) holds  independently of the field content of the operator (\ref{link}), in the following the  Stueckelberg field $\Phi$ will  only play the role of an auxiliary field, necessary for the  implementation of a continuous nonperturbative gauge invariant regularization.
Except for the gauge fixing term of the action (\ref{actionaphi}) which is proportional to $1/\xi$, the full action (\ref{action1}) is invariant under the following set of transformations 
\beqa
\label {gaugetrans1}
\ \psi(x)~&\rightarrow &~e^{i\int dy~\rho_3(x,y)\epsilon(y)}\psi(x)  \\
\label {gaugetrans2}
\ \Phi (x)  ~&\rightarrow &~  \Phi (x) + \frac{M}{e}\epsilon(x)  \\
\label {gaugetrans3}
\ \A ~&\rightarrow &~\A -\frac{1}{e}\d \epsilon(x),
\eeqa
which are a generalization of the  Pauli-Stueckelberg $U(1)$  gauge transformation \cite{PAUL}.
Here we must emphasize that the gauge invariance of the two first terms of the action (\ref{actionaphi}) do not impose any constraint for the scalar function $\epsilon$.
 
The dynamics of the  gauge field interacting with the fermion field is described in configuration space by the  path integral 
\beq
\label {zx}
\ Z_{Reg}(J)=\int \D\psi\D\bpsi\D A_{\mu}\D \Phi~e^{i\left(S_{Reg}(\psi ,\bpsi 
,A,\Phi)+\int 
dx~J_{\mu }(x)A^{\mu }(x)\right)},
\eeq
where  $ J_{\mu } $ is the source for the gauge field which we suppose to be conserved.
Since $\epsilon(x)$ is an arbitrary dimensionless scalar function, we choose now  $\epsilon (x)=e \Phi (x)/M$ and perform on the path integral the set of transformations (\ref{gaugetrans1}), (\ref{gaugetrans2}) and (\ref{gaugetrans3}), in this order.
Under this specific change of field variable the  transformed fermion field is gauge invariant and screened.
Physically this new field variable describes a state consisting of a bare electron charge surrounded by a cloud of massless $\Phi$  bosons which propagate the gauge transformation in a space-time  domain whose extension is defined through the function  $\rho_3$ entering the link operator (\ref{link}).
This picture is reminiscent  of a gauge invariant dressed electron field constructed in \cite{LAVELLE} in order to render the electron mass operator IR finite.

In the subset  of the fermionic and gauge field degrees of freedom this transformation is unitary, and the boson field $\Phi$ is rescaled by a factor $2$.
The Jacobian is that of a unitary transformation multiplied by an infinite numerical factor which can be absorbed in the normalization factor of the path integral.
It follows that only the action of the  transformed path integral is changed.
In terms of the new field variables the  action  (\ref{action1}) is given in momentum space by
\beqa
\label {action2}
\ S'_{Reg}(\psi,\bpsi,A,\Phi)&=&\int\dk ~\rho_1^2(k) \bpsi(k) \left(\kS-m 
\right)\psi(k)\nonumber \\
&& -e\int d\bar{p} d\bar{p'}~\rho_2(p)\rho_2(p')\rho_3(p-p')\bpsi(p)\left [\AS(p-p') -i(\pS -\pS') \frac{\Phi(p-p')}{M} \right ]\psi(p')\nonumber \\
&&+\int\dk ~ \left [\frac{1}{2} \left (k^2 -\frac {k^4}{ M^2} \right )\Phi(k)\Phi(-k) -ik_{\mu} \left (M - \frac {k^2}{ M} \right ) \Phi(k) A^{\mu}(-k) \right ]\nonumber \\
&& -\frac{1}{2} \int \dk ~  \left [\left( k^2 - M^2 \right )g_{\mu \nu } + \frac{1}{\xi} k_{\mu } k_{\nu }  \right] A^{\mu }(k) A^{\nu }(-k).
\eeqa
\mysection{QED as an effective regularized action}
If we integrate over the boson field $\Phi$ and add a source term for the transformed fermion field,  we obtain in momentum space the following regularized path integral
\beq
\label {zp}
\ Z(\eta,\bta,J)=\int \D\psi\D\bpsi\D A_{\mu}~e^{i\left(S(\psi ,\bpsi 
,A)+\int \dk ~ (\bta(k)\psi(k)+\bpsi(k)\eta(k)+J_{\mu }(-k)A^{\mu 
}(k))\right)},
\eeq
where $S(\psi ,\bpsi ,A)$ is the effective action given by
\beqa
\label {action3}
\ S(\psi,\bpsi,A)&=&\int\dk ~\rho_1^2(k) \bpsi(k) \left(\kS-m 
\right)\psi(k)\nonumber \\
& & -e\int d\bar{p} d\bar{p'}~\rho_2(p)\rho_2(p')\rho_3(p-p')\bpsi(p)A^{\mu}(p-p')\Gamma_{\mu }(p-p')\psi(p')\nonumber \\
& & -\frac{e^2}{2}\int\dk d\bar{p} d\bar{p'}~\frac{1}{k^2 (M^2-k^2)}\rho^2_3(k)\rho_2(p)\rho_2(k+p)\rho_2(k+p')\rho_2(p')\times \nonumber \\
& &\bpsi(p)\kS \psi(k+p) \bpsi(k+p')\kS \psi(p') \nonumber \\
& & -\frac{1}{2} \int \dk ~  \left [\left( k^2 g_{\mu \nu }  -  k_{\mu } k_{\nu } \right ) \left (1 -\frac{M^2}{k^2} \right ) + \frac{1}{\xi} k_{\mu } k_{\nu } \right] A^{\mu }(k) A^{\nu }(-k). 
\eeqa
In this expression  the matrix   $\Gamma_{\mu }(k)$ is defined by
\beq
\label{gam1}
\ \Gamma_{\mu }(k)=\gamma_{\mu }-\kS\frac{k_{\mu }}{k^2},
\eeq
and is transverse.
In the regularized action  (\ref{action3}), only the transverse part of the gauge field interacts with the fermion field and the photon mass is defined through the gauge  invariant mass term \cite{JLJ}
\beq
\label{gaugemass}
\frac{M^2}{2} \int \dk ~  \left( g_{\mu \nu }  -  \frac {k_{\mu } k_{\nu }}{ k^2} \right )  A^{\mu }(k) A^{\nu }(-k). 
\eeq
As a consequence,  this action  is  fully gauge invariant except for the gauge fixing term which is proportional to $1/\xi$.
This result is not surprising.
In the gauge invariant sector of the transformed action (\ref{action2}),  the longitudinal part of the gauge field is always compensated by the Stueckelberg field $\Phi$.
Therefore, integrating out the boson field is in some sense equivalent to integrate out the longitudinal part of the gauge field.
This explains why the four fermion vertex in (\ref{action3}) is singular; indeed it is the effective interaction between  the screened fermion field which is induced by the longitudinal part of the gauge field.
This can be verified directly from the action (\ref{action2}) by integrating out the  longitudinal modes of the gauge field in the unitary gauge i.e., $\xi = +\infty $ \cite{ZINN}.
As we will see later, this singularity is harmless when we calculate physical amplitudes, i.e., when we take into account all the degrees of freedom of the gauge field.

Since the part of the photon propagator induced by  (\ref{gaugemass})   decreases  like $1/k^4$ for high $k^2$, one can check  by power counting that the  operator (\ref{gaugemass})  does not induce any new UV divergences.
Moreover, since the fermion vertex is transverse only the IR regular  part $1/(k^2 -M^2)$ of the photon propagator contributes to the higher order loop corrections.
As a result the mass parameter $M$ is not renormalized by higher order radiative corrections but its value can be  fixed by the normalization conditions.

If one defines  the Legendre transform $\Gamma$ of the 1PI generating functional
\beq  
\label {gf}
\ G(\eta,\bta,J) \equiv \log Z(\eta,\bta,J),
\eeq
the invariance of the path integral (\ref{zp}) under the gauge field transformation (\ref{gaugetrans3}) is reflected by the Ward-Takahashi (WT) identity,
\beq
\label{WTindentity1}
q_{\mu}(2\pi)^4 \frac{\delta \Gamma }{A_{\mu }(q)}=-\frac{1}{\xi}q^2q_{\mu} A^{\mu }(-q),
\eeq
where the C-number function $A^{\mu }$ is the vacuum expectation value of the gauge field.
Despite the fact that the fermion field does not enter into the derivation of (\ref{WTindentity1}), we can show with the help of the DS equations of motion of the fermion field that the path integral (\ref{zp}) is, as expected, also invariant under the transformation (\ref{gaugetrans1}).
The identity  (\ref{WTindentity1}) implies that all $(2n+m)$-point functions constructed from $2n$ gauge fields and $m$ fermion fields are transverse with respect to the photon field for $n>1$ and $m \geq 0$.
Notice that this property holds true whether or not we consider off mass-shell or on mass-shell amplitudes, in contradistinction to  standard QED  where this  property is only true for pure photonic  $2n$-point functions when  we consider off mass-shell amplitudes.
What is the physical meaning  of the fermionic quartic interaction proportional to $e^2$ which are defined in  (\ref{action3}) ?

Since   the lowest order vertex function (\ref{gam1}) is transverse,  at first glance the currents associated to  external and internal photons lines are both algebraically conserved.
The need of the quartic fermionic interaction is there in order to ensure that  the transverse part of the photons will only contribute to the external lines, and not to the internal photons lines. 
This can be easily checked if one integrates out the photon field in the action  (\ref{action3}).
In this case the vertices associated to the lowest order effective quartic fermionic interaction
\beqa
\label{fermioneffaction1}
S_{Fermion}&=& -\frac{e^2}{2}\int\dk d\bar{p} d\bar{p'}~\frac{1}{M^2-k^2}\rho^2_3(k)\rho_2(p)\rho_2(k+p)\rho_2(k+p')\rho_2(p')\bpsi(p)\gamma_{\mu } \psi(k+p)\times \nonumber \\
& &  \bpsi(k+p')\gamma^{\mu } \psi(p'),
\eeqa
are the standard ones and are therefore clearly not  obtained only through the exchange of  transverse photon.
Notice that since the mass parameter $M^2$ in (\ref{fermioneffaction1}) is due to the  gauge invariant mass term  (\ref{gaugemass}), which as said before will play the role of counterterm,  this  parameter must be discarded in the lowest order calculation.

\begin{figure}
        \epsfysize=3cm
        \centerline{\hbox{\epsffile{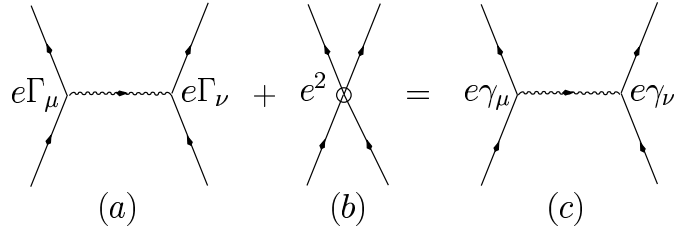}}}
\caption{The sum of the diagram (a) and  of the  diagram (b) associated to the quartic fermionic interaction defined in (\ref{action3}) gives the standard amplitude (c).
\label{tree_four}}
\end{figure}

This cancellation mechanism is reflected by the Feynman diagram of Fig.\ref{tree_four}.
As a consequence, the apparent  singularity of the four fermion vertex proportional to $1/k^2(M^2 -k^2)$ in (\ref{action3}) does not appear in any Feynman amplitude because it is algebraically cancelled by the singular part of the vertex function proportional to (\ref{gam1}).
In our previous work \cite{JLJ} the lowest order vertex was not transverse, but the regularized action contains an infinite sum of higher order vertices in the photon field in order  to render all diagrams with only external photon lines gauge invariant.

The  form of the effective actions  (\ref{action3}) and (\ref{fermioneffaction1}) implies that the Feynman diagrams have the following structure.
1)All vertices connected to internal photon lines are proportional to the standard Dirac matrices.
2)On the contrary all external photon lines are still  connected to the  matrices (\ref{gam1}).
However since the two polarization vectors $\epsilon_{\mu}(k)$  of the physical photon state of momentum $k$ are transverse and can always be defined in such a manner that $k^{\mu}\epsilon_{\mu}(k)=0$  \cite{WEIN}, it follows that $\epsilon_{\mu}(k)\Gamma^{\mu}(k)\equiv \epsilonS$ and consequently the vertices connected to the physical photon become   proportional to the  Dirac matrices.
As a result all the tree level Feynman amplitudes calculated from the regularized action (\ref{action3}) are those of standard QED.
Is this scenario the same if we take into account the radiative corrections ?

Since the lowest order vertex which is attached to the external photons lines is given exactly as in \cite{JLJ} by the transverse matrix (\ref{gam1}),  the expression of the  one loop order polarization operator will be that of  QED after subtraction of the unwanted quadratic divergence.
We will now show that the vanishing  of the renormalized photon mass will impose that $M^2$ must be a linear function of the square of the cutoff $\Lambda$.
Moreover we will  also confirm in the next section that the  one loop order expressions for the 1PI functions are exactly those expected from standard QED.
\mysection{The regularized DS equations}
If one wants to regularize the theory in the IR and UV domain only by the use of  cutoff functions, one can choose for (\ref{cutf2}) the following set
\begin{align}
\label{cutf3}
\rho_1(k)=1, \qquad \rho_2(\frac{k^2}{\Lambda^2})=e^{a_2\frac{ k^2}{\Lambda^2}}, \qquad  \rho_3(\frac{k^2}{\Lambda^2},\kappa)=\sqrt{\frac{k^2}{k^2 -\kappa^2}} e^{a_3\frac{ k^2}{\Lambda^2}} .
\end{align}
Here the $a_i$ are positive real numbers which parametrize the shape of these functions.
This is one of the  simplest choice in order to simplify the calculation of the  one loop order 1PI functions.
Notice that in Euclidean space  $ \rho_3$ is a real function and that it is only the square of  $ \rho_3$ which is associated with the photon propagator. 
Therefore, any  potential IR divergent term $1/k^2$ will be  replaced by its regularized form $1/(k^2-\kappa^2)$. 
 
\begin{figure}
        \epsfysize=2.3cm
        \centerline{\hbox{\epsffile{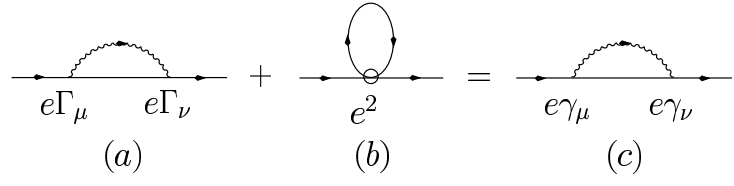}}}
\caption{The contribution of the diagram (b) which originates from the four fermions interaction cancels exactly the extra terms resulting from the structure of $\Gamma_{\mu }$ in the diagram (a).
\label{mass_op}}
\end{figure}

First we start with the polarization operator.
In terms of  the Legendre transform $\Gamma$  of the 1PI generating functional   (\ref{gf}) the relevant DS equation of motion is
\beqa
\label {equamotion1}
\ (2\pi)^4 \frac{\delta \Gamma }{\delta A^{\mu }(q )}&=&-\left[ \left( q^2g_{\mu \nu }  -q_{\mu } q_{\nu }\right)  \left( 1- \frac{M^2}{q^2}\right)+\frac {1}{\xi }q_{\mu } q_{\nu } \right ]   A^{\nu }(-q ) \nonumber \\
& &  +e(2\pi)^8 Tr\int d\bar{k} ~\rho_2(k)\rho_2(k-q)\rho_3(q) \frac{\delta^2G }{\delta \bta(k-q)\eta(k)}\Gamma_{\mu }(q).
\eeqa
If we take the functional derivative of  (\ref{equamotion1}) with respect to the vacuum expectation value of the field $A^{\nu}(-q')$, express the result in terms of the full electron propagator 
\beq
\label {S}
\ S(p',p)~=~-i(2\pi )^4\frac{\delta^2 G}{\delta \bta (p') \delta \eta (p)},
\eeq
and of the full electron photon vertex, 
\beq
\label {vertexone}
\  (2\pi)^8 \frac{\delta^3 \Gamma }{\delta A^{\mu }(-q )\delta \bpsi (p) 
\delta 
\psi 
(p')}~=~  \Gamma^{(3)}_{\mu}(p,p',q)~\equiv ~e\delta 
(p'-p-q)\Gamma^{(3)}_{\mu}(p,p')
\eeq
the use of translation invariance leads to 
\beq
\label {polarizationop1}
\Gamma^{(2)}_{\mu \nu} (q,q')= \delta (q-q') \Gamma^{(2)}_{\mu \nu} (q)
\eeq
with 
\beq
\label {polarizationop2}
\Gamma^{(2)}_{\mu \nu} (q)=- \left( q^2 g_{\mu \nu } -q_{\mu } q_{\nu } \right)   \left[ 1+\pi (q^2)- \frac{M^2}{q^2 }\right ] - \frac {1}{\xi }q_{\mu } q_{\nu },
\eeq
for the inverse of the exact photon propagator.
Here
\beq
\label {polarizationop3}
\pi(q^2)=-i\frac{e^2}{3q^2} Tr\int d\bar{k} ~\rho_2(k)\rho_2(k-q)\rho_3(q) S(k-q) \Gamma^{(3)}_{\mu}(k-q,k) S(k)\Gamma^{\mu }(q)
\eeq
contains all higher order corrections to   the polarization operator.
The transversality of the polarization operator (\ref{polarizationop2}) is a direct consequence of the WT identity (\ref{WTindentity1}).
At the one loop order  the vertex $\Gamma^{(3)}_{\mu}(k-q,k)$ is proportional to $\rho_3(q) \Gamma_{\mu}(q)$ and the expressions (\ref{polarizationop2}) and (\ref{polarizationop3}) are similar  to those found for the polarization operator in  \cite{JLJ}.
In addition to the standard regular piece which is logarithmic divergent, the expression for $\pi(q^2)$ contains   a  contribution  proportional to $1/q^2$ which shows a quadratic divergence.
This term which can give a mass to the photon was  removed in  \cite{JLJ} by the addition in the regularized action of a counterterm whose structure is exactly that of (\ref{gaugemass}).
The difference between the two schemes is that now the form of this term is dictated on the onset by the necessity of gauge invariance.
With the choice (\ref{cutf3}) of cutoff functions  and  when   the IR regulator mass  $\kappa$ is set to zero, the unwanted contribution to the photon mass is cancelled if we impose the renormalization condition
\beq
\label {renormcond1}
\left. q^2 \pi(q^2)\right |_{\stackrel {q^2=0}{}  } = M^2.
\eeq
Since the mass parameter $M^2$ (\ref{gaugemass}) is not renormalized by higher order radiative corrections its value is then  fixed to,
\beq
\label {Msquare1}
M^2=  e^2\Lambda^2 c,
\eeq
with
\beq
\label {Msquare2}
c=  \frac{1}{8\pi^2} \left ( \frac{1}{4a_2} -\frac{m^2}{\Lambda^2} \right ).
\eeq
The regularization in the IR domain can also be worked out by the introduction of a small photon mass  $\kappa$.
In this case the choice of  cutoff functions  can still be given by (\ref{cutf3}) but with $\rho_3(k)=1$.
Thus in this standard IR regularization method $M^2$ is given by 
\beq
\label {Msquare3}
M^2= \kappa^2 + e^2\Lambda^2 c.
\eeq
In order to deduce the mass operator and the vertex function we need the DS equation of motion which is associated with the invariance of the path integral under the translation of the fermion field, i.e.
\beqa
\label {equamotion2}
& &\left(\pS-m \right)^{ab}\psi^b(p) -(2\pi)^4 \frac{\delta \Gamma }{\delta \bpsi^a(p)}   -e\int  d\bar{k}~\rho_2(p)\rho_2(k)\rho_3(p-k)A^{\mu}(p-k)\Gamma_{\mu }^{ab}(p-k)\psi^b(k)\nonumber \\
& & + e (2\pi)^8 \int  d\bar{k}~\rho_2(p)\rho_2(k)\rho_3(p-k)\Gamma_{\mu }^{ab}(p-k)\frac{\delta^2 G}{\delta J_{\mu}(k-p)\delta \bta^b(k)} \nonumber \\
& & +ie^2 (2\pi)^{12} \int  d\bar{k} d\bar{k'} \frac{1}{k^2 (M^2-k^2)} \rho^2_3(k) \rho_2(k'-k)\rho_2(k')\rho_2(p+k)\rho_2(p)\left [\frac{\delta^3 G}{\delta \eta^c(k') \delta \bta^d(k'-k) \delta \bta^b(p+k) } \right. \nonumber \\
& & \left. -\frac{i}{(2\pi)^4}\frac{\delta^2 G}{\delta \eta^c(k')\delta \bta^b(p+k)} \psi^d(k'-k) + \frac{i}{(2\pi)^4}\frac{\delta^2 G}{\delta \eta^c(k')\delta \bta^d(k'-k)} \psi^b(p+k) \right ]\kS^{cd} \kS^{ab}\nonumber \\
& &  +\mathcal{O}(\psi \psi \bpsi)=0.
\eeqa
If we take the functional derivative of (\ref {equamotion2}) with respect to the C-number function $\psi^{b'}(p')$ and express the result in terms of the propagators and of the relevant 1PI functions, we obtain at the one loop order for the inverse of the electron propagator
\beq
\label {massoperator1}
 \Gamma^{(2)}(p)= \pS-m -\Sigma (p),
\eeq
where the mass operator $\Sigma$ is the sum of two terms
\beqa
\label {massoperator2}
\Sigma (p)&=& ie^2 \int  d\bar{k}~\rho_2^2(p)\rho_2^2(k+p)\rho_3^2(k) D^{\mu \nu}(k)\Gamma_{\mu }(k) S(k+p) \Gamma_{\nu }(k)  \nonumber \\
& & +ie^2\int  d\bar{k}~\rho_2^2(p)\rho_2^2(k+p)\rho_3^2(k) \frac{1}{k^2(M^2-k^2)} \kS S(k+p)\kS  +\mathcal{O}(e^4).
\eeqa
In this expression, $S(k)$ is the free electron propagator $1/(\kS-m)$ and $ D_{\mu \nu}(k)= g_{\mu \nu}/(M^2-k^2) + k_{\mu} k_{\nu} \cdots +\mathcal{O}(e^2)$ is the  expression of the photon propagator deduced from (\ref{polarizationop2}).
Moreover since $M$ is actually proportional to $e$ (\ref{Msquare1}), it is  assumed that at the end of the calculation only the zeroth order of the expansion of the inverse of  $(M^2-k^2)$ in power serie's of $e^2$  will be kept.
Since the vertices entering    the first term of the expression  (\ref{massoperator2}) are transverse, the longitudinal part of the photon propagator does not contribute, and the extra terms resulting from the structure of $\Gamma_{\mu }(k)$ (\ref{gam1}) are proportional to $ \kS S(k+p)\kS$.
As represented in Fig.\ref{mass_op} these extra  terms are  exactly cancelled by the last term of (\ref{massoperator2}), which originates from the four fermion interaction  defined in (\ref  {action3}).
As stated before, due to the quartic fermion interaction,  the effective vertex  connecting  the internal photon line to the  electrons lines is no longer transverse, and we find for $ \Sigma (p)$ at the one loop order the following IR and UV regularized expression
\beq
\label {massoperator3}
\Sigma (p)= - ie^2 \int  d\bar{k}~\rho_2^2(p)\rho_2^2(k+p)\rho_3^2(k) \frac{1}{k^2}\gamma_{\mu } S(k+p) \gamma^{\mu },
\eeq
which is the textbook result in the Feynman gauge up to the renormalization constant  \cite{JLJ}. 

\begin{figure}
        \epsfysize=3cm
        \centerline{\hbox{\epsffile{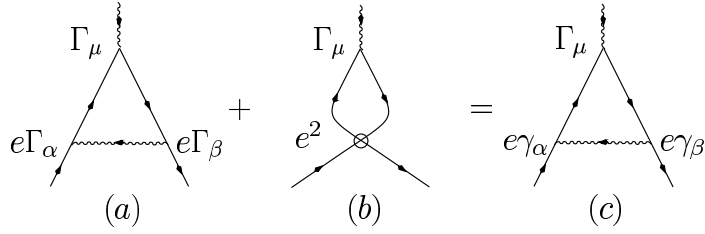}}}
\caption{Cancellation of the extra terms resulting from the structure of $\Gamma_{\mu }$ in the vertex function.
\label{vertex_op}}
\end{figure}

In the same manner  we obtain the expression of the vertex function  by taking the functional derivative of (\ref{equamotion2}) successively with respect to $\psi^{b'}(p')$ and $A^{\mu}(-q)$.
Up to the two loop order the vertex function is given by
\beqa
\label {vertexfunction2}
\Gamma^{(3)}_{\mu}(p,p')&=&\rho_2(p)\rho_2(p')\rho_3(q) \Gamma_{\mu }(q)+ i\frac{e^2}{(2\pi)^4}\int  dk~\rho_2^2(k+p)\rho_2^3(k+p')\rho_3^2(k) D^{\alpha \beta}(k)\times \nonumber \\
& & \Gamma_{\alpha}(k)S(k+p)\Gamma_{\mu }(q)S(k+p')\Gamma_{\beta}(k)   + i\frac{e^2}{(2\pi)^4}\int  dk~\rho_2^2(k+p)\rho_2^3(k+p') \rho_3^2(k)\times \nonumber \\
& &   \frac{1}{k^2(M^2-k^2)}\kS S(k+p)\Gamma_{\mu }(q)S(k+p') \kS   -i\frac{e^2}{(2\pi)^4}  \frac{\qS}{q^2(M^2-q^2)}\rho_3^2(q) \times \nonumber \\
& &
\int  dk~\rho_2^2(k+p)\rho_2^3(k+p')Tr \left [  S(k+p)\Gamma_{\mu }(q)S(k+p') \qS \right ] +\mathcal{O}(e^4),
\eeqa
$q=p'-p$ being the momentum transfer.
In this expression the second and last integral which are induced by the four fermion interaction  (\ref  {action3}) are represented by the diagram (b) of  Fig.\ref{vertex_op}.
As in the calculation of the mass operator, the second integral cancels exactly the contribution of the terms $ k_{\alpha}\kS /k^2$ and  $ k_{\beta}\kS/k^2$ coming from  the matrices (\ref{gam1}) $\Gamma_{\alpha}$ and $\Gamma_{\beta}$ of the first integral.
Moreover the last integral is in fact proportional to the evanescent operator $\int  dk~\rho_2^2(k)\rho_2^3(k-q) Tr\left [ \left (S(k-q)-S(k) \right) \Gamma_{\mu }(q) \right ]$ and behaves like $\mathcal{O}(\ln \Lambda/ \Lambda^2)$.
As a result the vertex function is given to the one loop order by
\beqa
\label {vertexfunction3}
\Gamma^{(3)}_{\mu}(p,p')&=& \Gamma_{\mu }(q)- i\frac{e^2}{(2\pi)^4}\int  dk~\rho_2^2(k+p)\rho_2^3(k+p')\rho_3^2(k) \frac {1}{k^2} \gamma_{\alpha}S(k+p)\Gamma_{\mu }(q)\times \nonumber \\
& & S(k+p')\gamma^{\alpha} .
\eeqa
Due to the WT identity  (\ref{WTindentity1}) the vertex function (\ref{vertexfunction3}) is transverse.
When the electron lines are on the mass-shell we can easily deduce  from the conservation and the Lorentz covariance of the one-particle matrix element of the current $\bu \Gamma^{(3)}_{\mu} u$  that the structure (\ref{gam1}) of the matrix $ \Gamma_{\mu }(q)$  implies that  the vertex function (\ref{vertexfunction3})  can be written as
\beq
\label {vertexfunction4}
\Gamma^{(3)}_{\mu}(p,p')= \Gamma_{\mu }(q)\left (1 +F_1\left (q^2 \right ) \right) + \frac{i}{2m} 
\sigma^{\mu \nu} q_{\nu}F_2(q^2),
\eeq 
where, as shown in   \cite{JLJ},  $F_1(q^2)$ and $F_2(q^2)$ are the standard form factors.
Notice that  we recover the textbook result of QED for the one-particle matrix element of (\ref{vertexfunction4}) when the electrons lines are on the mass-shell.
\mysection{Conclusion and outlook}
We have constructed an effective action by integrating out exactly the auxiliary Stueckelberg field of an underlying $U(1)$ gauge invariant and regularized theory whose fermionic part was described in  \cite{JLJ}.
The new action thus obtained,  is fully gauge invariant, regularized and describes the interaction of the photon with a massive screened fermion.
The predictions of QED are recovered at least at the tree level and to the one loop order.
Since the regularization is implemented non perturbatively at the level of the action through cutoff functions which are quite arbitrary, this regularization is well suited to probe or to elaborate exact RGE techniques for the study of abelian gauge theories.
It will be interesting to enlarge this construction to the case of chiral symmetry or to the case of non-abelian gauge symmetry.
\vskip 5mm
\centerline{\bf Acknowledgements}
I would like to thank J. Polonyi for very stimulating discussions.

\end{document}